\documentclass[oneside,a4paper,onecolumn,11pt]{article}

\usepackage{array}
\usepackage[caption=false,font=normalsize,labelfont=sf,textfont=sf]{subfig}
\usepackage{textcomp}
\usepackage{stfloats}
\usepackage{verbatim}
\usepackage[flushleft]{threeparttable}

\usepackage{fullpage}
\usepackage[left=2cm,top=2cm,bottom=2cm,right=2cm,includehead,nomarginpar,headheight=16pt]{geometry}
\usepackage[ruled, linesnumbered, vlined, commentsnumbered]{algorithm2e}
\usepackage{graphicx,subfig} 
\usepackage{amsfonts,amssymb,amsmath,amsthm,amsopn,mathtools}	
\usepackage{booktabs,diagbox,colortbl,multirow,tabularx,threeparttable,hhline}
\usepackage[listings,skins,breakable]{tcolorbox}

\usepackage{fancyhdr,fancyheadings,nopageno,lastpage} 
\usepackage{url}
\usepackage{enumerate}
\usepackage[shortlabels]{enumitem}
\usepackage{csquotes}
\usepackage{authblk}
\usepackage{footnote}
\usepackage{hyperref}
\usepackage{prettyref}
\usepackage{cite}
\usepackage{setspace}
\usepackage{color}
\usepackage{xcolor}  
\usepackage{geometry}
\usepackage{academicons}
\usepackage{fontawesome5}
\usepackage{pifont,ifsym,marvosym,manfnt} 

\usepackage{tikz}
\usepackage{pgfplots}
\usetikzlibrary{positioning,shapes,shadows,arrows,calc}
\tikzstyle{component}=[rectangle, draw=black, rounded corners, fill=blue!40, drop shadow, text centered, anchor=north, text=white, minimum height=1cm]
\tikzstyle{arrow}=[->, thick]

\pgfplotsset{compat=1.12}
\usetikzlibrary{intersections}
\usetikzlibrary{pgfplots.statistics}
\usepgfplotslibrary{fillbetween}

\fancypagestyle{plain}{%
    \fancyfoot[C]{}
    \fancyfoot[R]{\faLeanpub\ \, \thepage\ / \pageref{LastPage}}
    \fancyfoot[L]{\faBraille\ \textsc{COLALab Report} \faSlackHash\ \footnotesize\textifsym{2025}}
}

\hypersetup{
    colorlinks=true,
    linkcolor=ultramarine,
    filecolor=magenta,
    urlcolor=ultramarine,
    pdftitle={Overleaf Example},
    pdfpagemode=FullScreen,
}

\definecolor{red(munsell)}{rgb}{0.95, 0.0, 0.24}
\definecolor{navyblue}{RGB}{0, 0, 128}
\definecolor{myblue}{RGB}{34,31,217}
\definecolor{mycyan}{gray}{.7}
\definecolor{Gray}{gray}{0.9}
\definecolor{usccardinal}{rgb}{0.6, 0.0, 0.0}
\definecolor{ultramarine}{RGB}{0,32,96}
\definecolor{amber}{rgb}{1.0, 0.49, 0.0}

\newtheorem{remark}{\bf Remark}
\newtheorem{definition}{\bf Definition}[section]
\newtheorem{assumption}{\bf Assumption}[section]
\newtheorem{lemma}{\bf Lemma}[section]
\newtheorem{proposition}{\bf Proposition}[section]

\newcommand{\Real}{\mathbb R}
\newcommand{\norm}[1]{\left\Vert#1\right\Vert}

\newtcolorbox{quotebox}{colback=gray!10,boxrule=0.4pt,colframe=black,fonttitle=\bfseries,top=1pt,bottom=1pt}

\usepackage[ruled, linesnumbered, vlined, commentsnumbered]{algorithm2e}
\usepackage{prettyref}
\usepackage{amsfonts,amssymb,amsmath,amsthm,amsopn,mathrsfs,mathtools,wasysym}	
\usepackage{bm}
\usepackage{multirow}

\SetCommentSty{mycommfont}

\usepackage{lscape}

\newenvironment{code-example}
{
\vspace{0.15cm}
\noindent\begin{minipage}{\linewidth}
\begin{center}
\arrayrulecolor{black}
\color{black}
\begin{tabular}{|p{0.95\linewidth}|}
\hline%
\rowcolor{pink!20}%
}
{
\\\hline
\end{tabular}
\end{center}
\end{minipage}
\vspace{-0.2cm}
}

\begin{document}

\title{\vspace{-1ex}\LARGE\textbf{Optimal Parameter Adaptation for Safety-Critical Control via Safe Barrier Bayesian Optimization}~\footnote{This manuscript is under review (Preprint version, review only).}}

\author[1]{\normalsize Shengbo Wang}
\author[2]{\normalsize Ke Li}
\author[3]{\normalsize Zheng Yan}
\author[4]{\normalsize Zhenyuan Guo}
\author[5]{\normalsize Song Zhu} 
\author[6]{\normalsize Guanghui Wen}
\author[3]{\normalsize Shiping Wen}
\affil[1]{\normalsize School of Computer Science and Engineering, University of Electronic Science and Technology of China, Chengdu 611731, China}
\affil[2]{\normalsize Department of Computer Science, University of Exeter, EX4 4RN, Exeter, UK}
\affil[3]{\normalsize Australian AI Institute, Faculty of Engineering and Information Technology, University of Technology Sydney, NSW 2007, Australia}
\affil[4]{\normalsize College of Mathematics and Econometrics, Hunan University, Changsha, 410082, China}
\affil[5]{\normalsize School of Mathematics, China University of Mining and Technology, Xuzhou 221116, China}
\affil[6]{\normalsize Department of Systems Science, School of Mathematics, Southeast University, Nanjing, China}
\affil[\Faxmachine\ ]{\normalsize \texttt{shnbo.wang@foxmail.com}, \texttt{shiping.wen@uts.edu.au}}

\date{}
\maketitle

\vspace{-3ex}
{\normalsize\textbf{Abstract: } }Safety is of paramount importance in control systems to avoid costly risks and catastrophic damages. The control barrier function (CBF) method, a promising solution for safety-critical control, poses a new challenge of enhancing control performance due to its direct modification of original control design and the introduction of uncalibrated parameters. In this work, we shed light on the crucial role of configurable parameters in the CBF method for performance enhancement with a systematical categorization. Based on that, we propose a novel framework combining the CBF method with Bayesian optimization (BO) to optimize the safe control performance. Considering feasibility/safety-critical constraints, we develop a safe version of BO using the barrier-based interior method to efficiently search for promising feasible configurable parameters. Furthermore, we provide theoretical criteria of our framework regarding safety and optimality. An essential advantage of our framework lies in that it can work in model-agnostic environments, leaving sufficient flexibility in designing objective and constraint functions. Finally, simulation experiments on swing-up control and high-fidelity adaptive cruise control are conducted to demonstrate the effectiveness of our framework.

{\normalsize\textbf{Keywords: } }Safety-critical control, control barrier functions, Bayesian optimization, safe optimization.

\section{Introduction}
\label{sec:introduction}

In practical control scenarios, safety is of paramount importance during the design and implementation of control strategies. However, ensuring safety often comes at the expense of degrading control performance. Striking a good balance between maintaining satisfactory control performance and adhering to safety constraints remains an ongoing challenging task. In literature, there are three competitive frameworks for safe control, include receding-horizon-based model predictive control (MPC) \cite{ZhangMCHGL21} and reinforcement learning (RL) \cite{GarciaF15}, stability-inspired barrier Lyapunov function (BLF) \cite{TeeGT09,ChowNDG18} and control barrier function (CBF) methods \cite{AmesXGT17}, as well as reachability-based methods \cite{BansalCHT17}. Among these, the CBF method is becoming a promising solution for safe control due to its high computational efficiency and versatility for easy integration with other control modules \cite{Shaw-CortezOMC21}. Unfortunately, as also a conservative solution \cite{WangLWSZH23}, the CBF method involves various configurable parameters that are \emph{implicitly} related to both control performance and system safety \cite{LiWGZHSWW24}. In this work, we aim to develop a framework for parameter adaptation for CBF-based safety-critical control tasks (SC2Ts), focusing on safely and efficiently optimizing the safe control performance.

There are two main issues hindering the utilization of existing optimization techniques, such as evolutionary algorithms \cite{ArnoldH12} and derivative-based programming \cite{ParwanaP22,MaZTS22}, to calibrate the configurable parameters in SC2Ts. The first issue lies in the optimization efficiency under a black-box environment. The numerical control performance, such as the linear quadratic regulator (LQR) performance \cite{WangWSZH21}, may not be analytically dependent on the configurable parameters. However, measuring control performance can be computationally expensive, especially for high-fidelity simulations or real systems. As a result, we can only leverage zeroth-order and sample-efficient techniques for optimization \cite{WangL24}. The second issue is the feasibility of candidate configurations in safety-critical environments. Since a poorly calibrated configuration can lead to violations of safety constraints or pose higher risks, the trial-and-error approaches commonly used in safe RL are not suitable for SC2Ts. To address these challenges, we plan to leverage Bayesian optimization (BO), a recognized query-efficient framework for black-box optimization \cite{WangL24}. Furthermore, we will propose a safe version of BO that can explicitly incorporate feasibility-critical constraints during the parameter adaptation processes. While existing research on safe BO methods relies on the assumptions of Lipschitz-continuity of constraints \cite{BerkenkampKS16}, the proposed safe BO do not require these rigid assumptions.

Although considerable research has been conducted on parameter calibration and performance enhancement of control systems \cite{Martinez-Cantin19,MarcoBKHRT21,ZhuPB22,CosnerTTLMUAOYA22}, this paper proposes a principled performance optimization framework tailored for SC2Ts.

\begin{enumerate}
    \item We highlight the role of configurable parameters in CBF-based SC2Ts, and introduce a high-level structure of parameter adaptation as a constrained black-box optimization problem. For the first time, we provide a systematic categorization of involved parameters, offering valuable guidance to practitioners in parameter selection.
    \item For efficient and safe parameter adaptation, we propose a novel safe BO method that combines the barrier-based interior-point method and pessimistic bounds of Gaussian processes (GPs) to ensure feasibility during optimization. We also give theoretical criteria for feasibility and optimality of this method.
    \item We develop a novel framework that integrates the lower-level CBF method and the high-level safe barrier BO for optimal parameter adaptation. To illustrate the efficacy of this framework, we present two toy examples including tracking with obstacle avoidance and solving constrained LQR. Additionally, we conduct simulation experiments on swing-up control and adaptive cruise control (ACC) tasks to demonstrate the effectiveness.
\end{enumerate}

\textit{Notations}: Throughout this paper, $\Real$ denotes the real value space, while $\Real^n$ and $\Real^n_+$ denote $n-$dimensional real space and positive real space, respectively. For vector $x$ and matrix $A$, $\norm{x}$ and $\norm{A}$ are the Euclidean norm and induced matrix norm, respectively. 
For $x\in \Real^n$ and a continuously differentiable function $h(x): \Real^n \to \Real$, we denote its Lie derivative regarding $f(x)\in \Real^n$ as $L_h f(x) = \frac{\partial h } {\partial x}^{\top} f(x)$. For a compact set $\mathcal{C}$, $\text{Int}(\mathcal{C})$ and $\partial \mathcal{C}$ denote the interior and boundary of $\mathcal{C}$.

\section{Preliminaries}
\label{section:preliminaries}

\subsection{Problem Formulation}
\label{section:problem_formulation}

Consider the following autonomous systems:
\begin{equation}
    \dot x = f(x) + g(x) u, 
    \label{eqn:system_dynamics}
\end{equation}
where $x \in \mathcal{X}\subset \Real^n $ and $u \in \mathcal{U} \subset\Real^m$ are the system state and control input respectively, $f(x):\Real^{n} \to \Real^n$ denotes the drift dynamics, and $g(x):\Real^n \to \Real^{n \times m}$ is the input dynamics. Both $f$ and $g$ are locally Lipschitz continuous. The initial condition of \eqref{eqn:system_dynamics} is $x_0 = x(t_0)\in \mathcal{X}$ for some starting time $t_0$. For autonomous systems, we simplify the notation to $x$ and $u$ instead of $x(t)$ and $u(t)$.
In practical control scenarios, both system state and input should be constrained in task-specific spaces. Specifically, any violation of state constraints, i.e. $x \notin \mathcal{X}$, may lead to a catastrophic damage. In addition, we consider $\mathcal{U} = \left\{u\in \Real^m\vert A_u u\leq b_u \right\}$. Consequently, the safety-critical control tasks (SC2Ts) aims to synthesize a controller from $\mathcal{U}$ that enforces $x\in \mathcal{X}$ during control processes.

In this work, we are going to propose a high-level optimization framework for SC2Ts, which safely and efficiently optimizes the configurable parameters in SC2Ts to improve the practical performance across different scenarios. This is formulated as a constrained optimization problem:
\begin{equation}
        \mathrm{minimize}_{\mathbf{z} \in \Omega} \;  r(\mathbf{z}) \quad 
        \mathrm{s.t.} \quad \vec{g}(\mathbf{z})\leq 0, 
    \label{eq:cop}
\end{equation}
where $\mathbf{z}=(z_1,\cdots,z_p)^\top\in\Omega$ denotes the decision variable, i.e., the configurable parameters in SC2Ts, $\Omega=[z_i^\mathrm{L},z_i^\mathrm{U}]_{i=1}^p\subset\mathbb{R}^p$ is the search space, $z_i^\mathrm{L}$ and $z_i^\mathrm{U}$ are the lower and upper bounds of $z_i$ respectively. The objective function $r(\mathbf{z})$ and $q$ constraint functions $\vec{g}(\mathbf{z})=(g_1(\mathbf{z}),\cdots,g_q(\mathbf{z}))^\top$ 
are: $i)$ \emph{analytically unknown}, we do not have access to the formulation of $r$ or $\vec{g}$, but are given $N$ observations $\langle \mathbf{z}^i, r(\mathbf{z}^i), \vec{g}(\mathbf{z}^i) \rangle_{i=1}^N$; $ii)$ \emph{computational expensive}, the number of observations $N$ is limited; and $iii)$ \emph{feasibility-critical}, we should ensure that candidate solutions to \eqref{eq:cop} are feasible with high probability.
Under these harsh requirements, existing methods including differentiable convex programming (DOP), safe RL, and constrained BO, may not work well, as illustrated in Table \ref{tab:methodologycomparison}.

\begin{table}[tb]
    \centering
    \caption{The comparison of different optimization techniques}
    \begin{tabular}{c|c|c|c}
       Methods & Black-box & Efficiency & Feasibility \\
        \hline
       DOP \cite{ParwanaP22,MaZTS22} & $\times$ & - & $\checkmark$\\
       \hline
       Safe RL \cite{GarciaF15,ChowNDG18} & $\checkmark$ & $\times$ & - \\
       \hline
       Constrained BO \cite{CosnerTTLMUAOYA22,KhosraviMMSLR22}& $\checkmark$ & $\checkmark$ & $\times$ \\
       \hline
       Safe BO \cite{BerkenkampKS16, KrishnamoorthyD22} & $\checkmark$ & $\checkmark$ & $\checkmark$ \\
    \end{tabular}
    
    \label{tab:methodologycomparison}
\end{table}


\subsection{Control Barrier Function Method}
\label{section:cbf}
The CBF method paves a way to effectively  mediate the control objective and safety requirements when synthesizing a safe controller. We present a brief review of the CBF method for SC2Ts as follows, and the involved configurable parameters will be discussed later.
\begin{definition}[Control Barrier Function \cite{AmesXGT17}]
\label{definition:cbf}
For a continuously differentiable function $h(x):\Real^{n} \to \Real$ and a compact set $\mathcal{C}= \left\{ x\in \mathcal{X} \vert h(x)\ge0 \right\}$ with $h(x)>0, \forall x \in \text{Int}(\mathcal{C})$, and $h(x)=0, \forall x\in \partial \mathcal{C}$, $h(x)$ is called a (zeroing) CBF defined on $\mathcal{X}$ for systems \eqref{eqn:system_dynamics} with a relative degree $1$ if there exists an extended class $\mathcal{K}$ function $\bar\alpha$ such that  $\forall x \in \mathcal{X}$,
\begin{equation}
    \sup_{u \in \mathcal{U}}\Big( L_f h(x) + L_g h(x) u  \Big) \ge - \bar\alpha (h(x)). \label{eqn:definition_CBF}
\end{equation}
\end{definition}


A practical implementation of the CBF method is based on quadratic programming (QP), where the values of control input are computed in real-time by enforcing the safety requirements in \eqref{eqn:definition_CBF} and integrating with nominal controllers, formulated by
\begin{gather}
    u_{\text{safe}} = \arg\min_{u\in \Real^{m}} l_{\vartheta}(u,x) \label{eqn:cbfQP}\\
    \text{s.t.}~ L_f h(x) + L_g h(x) u  + \alpha h(x) \ge 0,\tag{\ref{eqn:cbfQP}{a}} \label{eqn:cbfQP_cbf}\\
    A_u u\leq b_u \tag{\ref{eqn:cbfQP}{b}} \label{eqn:cbfQP_saturation}.
\end{gather}
Herein, the objective function $l: \mathcal{U} \times \mathcal{X} \to \Real_+$ denotes the myopic loss for mediating control performance, the subscript $\vartheta$ denotes the configurable variable in the design of $l_{\vartheta}(u,x)$. For instance, given a nominal controller $k_{\mathrm{ref}}(x)$, the loss can be defined by $l_{\vartheta}(u,x) = \norm{u - k_{\mathrm{ref}}(x)}^2 + \lambda \norm{u}^2$ with $\lambda\ge 0$ the configurable parameter for regularization. Note that the choice of function $\bar\alpha$ is also configurable. For brevity, we stipulate $\bar\alpha(h(x)) = \alpha h(x)$ with $\alpha > 0$ to be determined.

\begin{remark}
    When a system has a relative degree more than one, i.e., $L_g h(x) = 0$, the exponential CBF can be introduced for a more conservative design of safe controllers \cite{NguyenS16,XiaoB22}. Similar to \eqref{eqn:cbfQP_cbf}, the exponential CBF method will introduce additional parameters \cite{NguyenS16} or functions \cite{XiaoB22} that need to be determined, thereby increasing the burden of calibration.
\end{remark}

\subsection{Gaussian Processes and Bayesian Optimization}

BO serves as a query-efficient framework for black-box optimization, consisting of surrogate modeling for unknown objectives and constraints and acquisition functions for efficient search. The most commonly used surrogate models are Gaussian processes (GPs) \cite{GPML}. Given the dataset $\mathcal{D}_r=\langle\mathbf{z}^i,r(\mathbf{z}^i)\rangle_{i=1}^{N}$ of the objective function in \eqref{eq:cop}, a GP model with mean $m(\mathbf{z})$ and noise-free likelihood predicts a distribution $\tilde{r}(\tilde{\mathbf{z}}) \sim \mathcal{N}(\mu_r(\tilde{\mathbf{z}}),  \sigma^2_r\left(\tilde{\mathbf{z}}\right))$ of a candidate solution $\tilde{\mathbf{z}}\in\Omega$ by:
\begin{equation}
    \begin{split}
        \mu_r(\tilde{\mathbf{z}})&= m(\tilde{\mathbf{z}}) + {\mathbf{k}^\ast}^\top K^{-1} \mathbf{r}^\prime,\\
        \sigma^2_r\left(\tilde{\mathbf{z}}\right)&=k(\tilde{\mathbf{z}},\tilde{\mathbf{z}})-{\mathbf{k}^\ast}^\top K^{-1} {\mathbf{k}^\ast},
    \end{split}
    \label{eq:GP}
\end{equation}
in which $\mathbf{k}^\ast$ is the covariance matrix between $Z$ and $\tilde{\mathbf{z}}$, $K$ is the covariance matrix of $Z$, $Z=\left(\mathbf{z}^1,\cdots,\mathbf{z}^N\right)^\top$ and $\mathbf{r}=\left(r(\mathbf{z}^1) - m(\mathbf{z}^1),\cdots,r(\mathbf{z}^N) - m(\mathbf{z}^N)\right)^\top$. In this paper, we use a trainable constant mean function and the Mat\'ern ${5/2}$ as the kernel function \cite{GPML}. The $i$-th constraint in \eqref{eq:cop} will be modeled by an independent GP $\tilde{g}_i(\tilde{\mathbf{z}})$ whose predictive mean and variance are denoted by $\mu_g^i(\tilde{\mathbf{z}})$ and $\sigma^i_g(\tilde{\mathbf{z}})$, respectively.

The search process of BO is driven by an acquisition function that automatically strikes a balance between exploiting the predicted optimum and exploring the uncertain areas that potentially contain better solutions. One of the most commonly used acquisition functions is the lower confidence bound (LCB) \cite{SrinivasKKS10}, which takes the following form:
\begin{equation}
    \mathrm{LCB}(\tilde{\mathbf{z}} \vert  \mathcal{D}_r) = \mu_r(\tilde{\mathbf{z}}) - \beta_r{ \sigma_r (\tilde{\mathbf{z}})}.
    \label{eqn:lcb}
\end{equation}
Herein, $\beta_r>0$ controls the inclination of exploration during search. The constrained version of BO is referred to \cite{WangL24,BerkenkampKS16}.

\section{Main Results}
\begin{figure}[tb]
    \centering
    \scalebox{.3}{\includegraphics{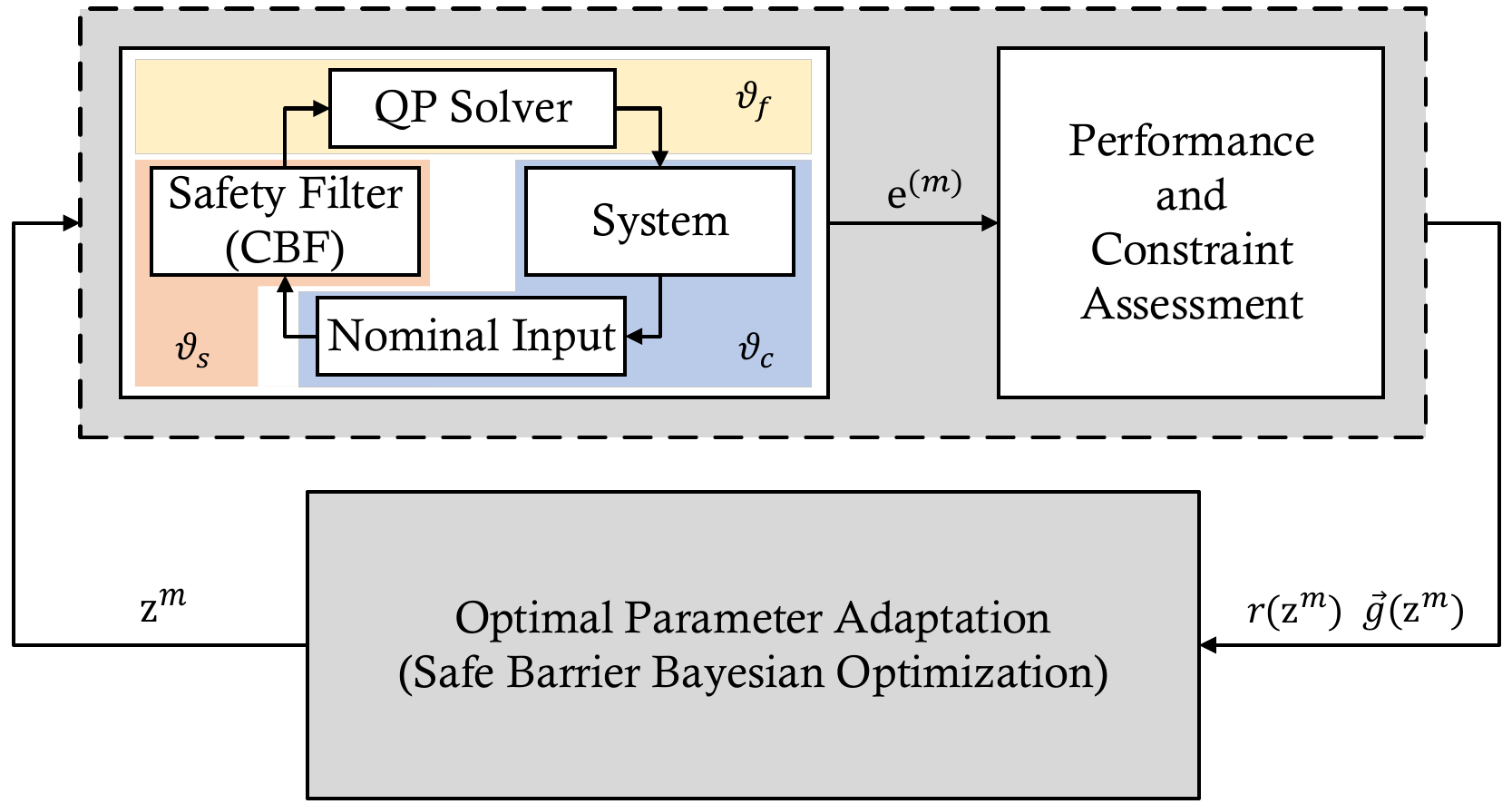}}
    \caption{An overview of the \texttt{CBF-SB2O} framework for SC2Ts.}
    \label{fig:structure}
\end{figure}

In this section, we will develop a safe and efficient parameter adaptation framework for SC2Ts. Firstly, we study the relationship between configurable parameters and control performance in SC2Ts, meanwhile presenting a systematical categorization of the involved parameters. Secondly, we propose a novel safe barrier Bayesian optimization (SB2O) method that ensures feasibility with high probability during search. Finally, we design an integrated framework of parameter adaptation for SC2Ts, dubbed \texttt{CBF-SB2O}, with detailed investigations. The overall framework is depicted in Fig. \ref{fig:structure}.

\subsection{Performance and Configurable Parameters of SC2Ts}
\label{section:performancedescription}
\subsubsection{Relationship between Performance and Configurations} 
The control performance of SC2Ts defined by reward functions in \eqref{eq:cop} is dependent not only on the configurable parameters but also dependent on $x$, $u$, and some inherent parameters $\mathbf{k}$, thereby we have $r(\mathbf{z}, \mathbf{k}, x, u)$. As an important example, the analytical format of the LQR performance \cite{GarciaF15} is
\begin{equation}
    \label{eqn:LQR_performance}
    r(\mathbf{z}, \mathbf{k}, x, u) = \int_{t_0}^{\infty} \left( x^\top Q x + u^\top R u \right)\; dt,
\end{equation}
where $Q$ and $R$ are predefined matrix parameters, i.e., $\mathbf{k}$. For a linear system, determining the optimal controller to \eqref{eqn:LQR_performance} is equivalent to computing the optimal coefficients of the linear feedback gain matrix, i.e. $\mathbf{z}$. For a nonlinear system, the performance can be optimized through adaptive dynamic programming (ADP) \cite{0005ZM021,WeiYSW22}. For both cases, the configurable parameters play a critical role in enhancing the performance. To get rid of the influence of $x$ and $u$, we study autonomous systems with fixed initial state $x_0$.


In what follows, we present an illustrative example to show the influence of configurations on the control performance.

\begin{figure}[tb]
    \centering
    \scalebox{.5}{\includegraphics{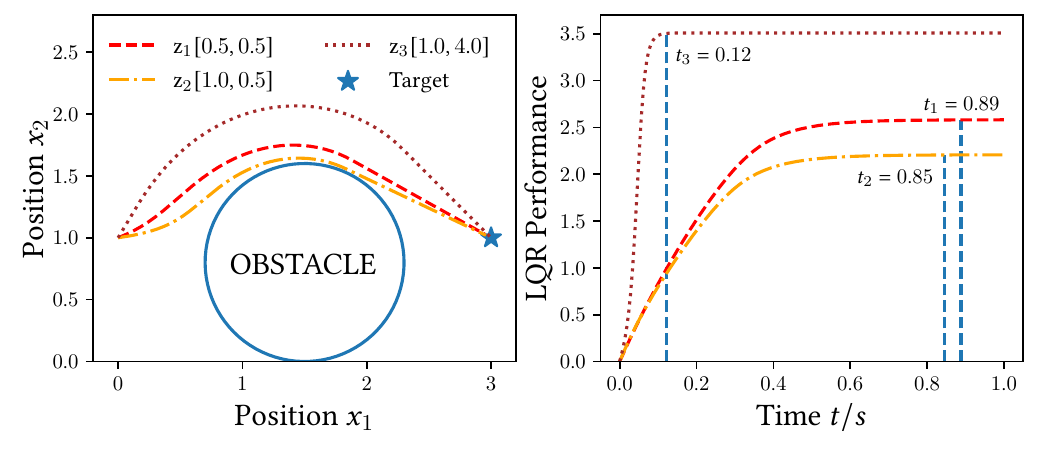}}
    \caption{A safe tracking control task with three different configurations. (Left) The moving trajectories with different $\mathbf{z}_i$. (Right) the LQR performance ($Q=I$ and $R=2I$) and the response time $t_i = \inf_{t>t_0} \{\norm{x(t)-x_d} \leq 0.1\}$.}
    \label{fig:toyexample}
\end{figure}

\emph{Illustrative Example 1}: Consider a tracking control task with obstacle avoidance requirements. A 2-D moving system, described by $\dot x = u$ with position $x$ and velocity input $u$, starts from $x_0 = [0,1]^\top$ and moves to a target position $x_d = [3, 1]^\top$. For tracking control, a nominal controller is designed by $k_{\text{ref}}(x) = k_p(x_d - x)$ with a coefficient $k_p>0$. There is an obstacle on the way, marked by a circle centered at $x_c = [1.5, 0.8]^\top$ with radius $r = 0.8$. By setting $h(x) = \norm{x - x_c} - r$, we obtained a valid CBF constraint as \eqref{eqn:cbfQP_cbf} with any $\alpha>0$. Additionally, we set $l_{\vartheta}(u,x) = \norm{u -k_{\text{ref}}(x)}^2$ in the QP \eqref{eqn:cbfQP} for a basic implementation of the CBF method \cite{WangWYCSH23}. Overall, the involved configurable parameters are $\mathbf{z} = \left[ \alpha, k_p \right]$ with $\Omega = \Real_+^2$. In Fig. \ref{fig:toyexample}, three runs were conducted with different $\mathbf{z}$, all tracked successfully without collision. However, the ranking of the three configurations varies according to different definitions of $r(\mathbf{z})$ in \eqref{eq:cop}, as demonstrated below.
\begin{itemize}
    \item Considering the LQR performance \eqref{eqn:LQR_performance}, $\mathbf{z}_2$ should be the best configuration among three. We may conclude that a larger $\alpha$ can lead to a better performance in this scenario.
    \item In terms of response time, $\mathbf{z}_3$ is a better choice since a larger $k_p$ results in faster convergence. If a response time $t>0.85\mathrm{s}$ is forbidden, $\mathbf{z}_1$ is infeasible and should not be tested in feasibility-critic environment \eqref{eq:cop}.
    \item For conservativeness, $\mathbf{z}_2$ is considered to be less conservative as it allows closer distance to the obstacle \cite{LopezSH21}, while $\mathbf{z}_1$ and $\mathbf{z}_3$ maintain a higher safety margin.
\end{itemize}

In summary, the control performance of a SC2T is highly related to the configurable parameters, and the optimal configuration vary according to different objectives \cite{HuangWL24}. Furthermore, most objectives and constraints, such as latency and robustness, may not have analytical formats. In general, we consider problem \eqref{eq:cop} a black-box optimization problem. Since simulations and physical experiments can be costly, the optimization problem is expensive to solve.

\subsubsection{Categorization of configurable parameters in SC2Ts} 
Based upon the above discussions, we provide a systematic categorization of configurable parameters in CBF-based SC2Ts, consisting of three aspects as follows.
\begin{itemize}
    \item \underline{\emph{Control parameters $\vartheta_c$}} related to specific control tasks. Although the CBF method enables a decoupled design of safety and control objective, for a well-calibrated nominal controller, the mediation between safety and control objective may degrade its performance and make it sub-optimal for SC2Ts. Therefore, $\vartheta_c$ contains the control parameters such as the proportional-integral-derivative (PID) \cite{KhosraviMMSLR22} coefficients and the linear matrix inequality (LMI) gains \cite{Khalil02}. Besides, implicit control parameters such as that in control Lyapunov function (CLF) \cite{AmesXGT17} are included in this set as well. 
    
    \item \underline{\emph{Safety parameters $\vartheta_s$}} in CBF method. The real-time control synthesis framework \eqref{eqn:cbfQP} necessitates the design of both the objective function, such as the regularization and balancing among control objectives, and the safety constraints, such as $\alpha$ in \eqref{eqn:cbfQP_cbf} and additional parameters in adaptive and robust CBF methods \cite{DeanTCRA20, CosnerTTLMUAOYA22}. Despite knowledge on the roles of some parameters, their impact towards control performance are rarely studied and can be significantly different across scenarios.
    
    \item \underline{\emph{Deployment parameters $\vartheta_f$}} from practical algorithm implementation. To reduce the real-time computational burden, numerical solutions to \eqref{eqn:cbfQP} are required, such as discretization and approximation, which will in turn lead to the performance degradation and even unsafe behaviors. These parameters include configurations of a specific QP solver such as the error tolerance and learning rate \cite{GillMW81,FazlyabPPR18}, and alternatives numerical solutions to \eqref{eqn:cbfQP} such as gradient-flow based method \cite{WangWYCSH23}. 
\end{itemize}

In all, the decision variable $\mathbf{z}$ in \eqref{eq:cop} is a subset of the union group $\Pi = \left\{\vartheta_c, \vartheta_s, \vartheta_f \right\}$. Solving \eqref{eq:cop} at a high level requires low level evaluation for the performance measurement and constraint violations. While a re-calibration in this way has the potential to obtain good control performance, it poses a risk of unexpected output such as  high evaluation costs and even unsafe behaviors. To this end, we introduce feasibility-critical constraints in \eqref{eq:cop} to circumvent this issue as much as possible, which will be studied in the next section.

\begin{remark}
    The criteria and principles for designing $\vartheta_c$ and $\vartheta_s$, such as LMI, can be further utilized as a feasibility guidance when searching for a better solution $\mathbf{z}$. For instance, policy search can be used to find LQR solution under semi-definite programming \cite{BalakrishnanV03,Fazel0KM18}. In effect, it narrows the configuration space $\Omega$, thereby improves optimization efficiency.
\end{remark}

\subsection{Safe Barrier Bayesian Optimization}

To cope with unknown feasibility-critical constraints, we propose a safe version of BO that ensures candidate solutions feasible with high probability. We make the following assumptions commonly used in BO and constrained optimization.

\begin{assumption}
The objective $r$ and all constraints in $\vec{g}$ are independent from each other. In addition, the $i$-th constraint, $g_i: \Omega \to \Real$, is in a Reproducing Kernel Hilbert Space (RKHS) of the kernel $k$ and have a bounded RKHS norm $\norm{g_i}_k$.
\label{assumption:rkhs}
\end{assumption}
\begin{lemma}[Calibrated uncertainty \cite{SrinivasKKS10}]
    Let $\delta \in (0, 1)$. If Assumption \ref{assumption:rkhs} holds, then the following inequality holds 
    \begin{equation}
        \vert \mu_g^i(\mathbf{\tilde z}) - g_i(\mathbf{\tilde z}) \vert \leq \bar \beta_{N+1}^i \sigma_g^i( \mathbf{\tilde z})
    \end{equation}
    with probability at least $1-\delta$ for all $\mathbf{\tilde z} \in \Omega$, where
    \begin{equation}
        \bar \beta_N^i = \left(2\norm{g_i}^2_k + 300 \gamma_{N}^i \ln^3(\frac{N}{\delta})\right)^{0.5},
    \end{equation}
    and $\gamma_{N}^i$ is the maximum information gain with $\langle \mathbf{z}^j, g_i(\mathbf{z}^j) \rangle^N_{j=1}$.
\end{lemma}

Based upon the above lemma, to avoid evaluating an infeasible configuration, we introduce a pessimistic estimation of each constraint and integrate it into \eqref{eqn:lcb} as follows:
\begin{equation}
\begin{split}
    \mathrm{minimize}&_{\mathbf{\tilde z} \in \Omega}  
    \mathrm{LCB}(\tilde{\mathbf{z}} \vert  \mathcal{D}_r) \\
    \mathrm{s.t.} \quad \mu_g^i(\tilde{\mathbf{z}}) - \bar \beta^i_N & { \sigma_g^i (\tilde{\mathbf{z}})} \leq 0, \quad i=1,\dots q.
\end{split}
\label{eq:constrainedSafeBO}
\end{equation}

\begin{proposition}
    \label{proposition:safety}
    Let $\delta \in (0, 1)$. If Assumption \ref{assumption:rkhs} holds, the optimal solution $\mathbf{\tilde {z}}^*$ to \eqref{eq:constrainedSafeBO} is safe/feasible in original problem \eqref{eq:cop} with probability at least $(1 - \delta)^q$.
\end{proposition}

The above result can be proven from the joint probability of independent events. In essence, solutions to \eqref{eq:constrainedSafeBO} inherit the merits of LCB for efficient search and ensure feasibility against uncertainty. Despite mathematically appealing, it is indeed another intractable nonlinear constrained optimization problem, different from \eqref{eq:cop}, with known objective and constraints. In order to mitigate the intractability, we modify \eqref{eq:constrainedSafeBO} and propose a barrier-based acquisition function as
\begin{equation}
    \mathrm{SB2O}(\tilde{\mathbf{z}} \vert  \mathcal{D}_{r,g})  =  \mathrm{LCB}(\tilde{\mathbf{z}} \vert  \mathcal{D}_r) + \sum_{i=1}^q \phi_g^i(\mathbf{\tilde z}),
\label{eq:SB2O}
\end{equation}
where $\phi_g^i(\mathbf{\tilde z}) = - c^{-1} \log \left(-\mu_g^i(\tilde{\mathbf{z}}) + \bar \beta^i_N  { \sigma_g^i (\tilde{\mathbf{z}})}\right)$ with $c>0$, and $\mathcal{D}_{r,g}=\langle\mathbf{z}^i,r(\mathbf{z}^i), \vec{g}_(\mathbf{z}^i)\rangle_{i=1}^{N}$ denotes the collection of $N$ evaluated objective and constraint pairs at the $(N+1)$th round. Noted that solving \eqref{eq:SB2O} is as trivial as solving the original LCB problem \eqref{eqn:lcb}. and the GP models enable us to leverage gradients for more efficient search. We quantify the feasibility and suboptimality of SB2O as follows. 

\begin{proposition}
    \label{proposition:sb2o}
    Let $\delta \in (0, 1)$ and the Assumption \ref{assumption:rkhs} hold. Assume also that the safe BO starts from a feasible point. At each following round, the optimal candidate solution $\mathbf{\tilde {z}}^*$ to \eqref{eq:SB2O} is safe/feasible with probability at least $(1 - \delta)^q$. Moreover, the solution is $q/c$-suboptimal as
    \begin{equation}
        \vert \mathrm{LCB}(\tilde{\mathbf{z}}^* \vert  \mathcal{D}_r) - \mathrm{SB2O}(\tilde{\mathbf{z}}^* \vert  \mathcal{D}_{r,g})  \vert \leq \frac{q}{c}. \label{eqn:prop_optimal}
    \end{equation}
\end{proposition}

\textit{Proof}. We will leverage the theoretical results from the interior-point method \cite{convex_book} to show the feasibility and suboptimality of the candidate solutions obtained by solving \eqref{eq:SB2O}. First, we introduce an equivalent transformation of \eqref{eq:constrainedSafeBO} as
\begin{equation}
    \mathrm{minimize}_{\mathbf{\tilde z} \in \Omega}  
    \mathrm{LCB}(\tilde{\mathbf{z}} \vert  D_r) + \sum_{i=1}^q I_{-}\left(\mu_g^i(\tilde{\mathbf{z}}) - \bar \beta^i_N  { \sigma_g^i (\tilde{\mathbf{z}})}\right)
\label{eq:indicatorCBO}
\end{equation}
in which $I_{-}:\Real \to\Real$ denotes the indicator function satisfying
\begin{equation*}
    I_{-}(a) = \left\{\begin{array}{cc}
         0,& a\leq 0, \\
         +\infty, & a>0.
    \end{array}\right.
\end{equation*}
Based on Proposition \ref{proposition:safety}, with probability at least $(1-\delta)^q$, the optimal solutions to \eqref{eq:indicatorCBO} is feasible regarding the original optimization problem \eqref{eq:cop}. To mitigate the intractability of the non-differentiable indicator function, a logarithmic approximation is used as $\phi(a)= - c^{-1} \log \left(-a\right)$, where $c>0$ determines the accuracy of the approximation, revealed by the fact that $\lim_{c\to\ + \infty} \phi(a)  = I_{-}(a), \forall a \in \Real$. As a result, \eqref{eq:SB2O} leverages this transformation for a differentiable approximation. Note that there should be a known feasible solution at the beginning to ensure this optimization problem is well-defined. As long as the initial search point is feasible, the candidate solutions to \eqref{eq:SB2O} is also feasible to \eqref{eq:indicatorCBO}, therefore with probability at least $(1-\delta)^q$ feasible to \eqref{eq:cop} in the context of SC2Ts.

As a compromise, the approximation introduces a bias from optima. We see the fact that, with any gradient-based optimizer \cite{GPML}, all solutions $\mathbf{\tilde {z}}^*$ to \eqref{eq:SB2O} have zero value on their gradients. Denoting $\tilde g_i(\mathbf{\tilde {z}}) = -\mu_g^i(\tilde{\mathbf{z}}) + \bar \beta^i_N  { \sigma_g^i (\tilde{\mathbf{z}})}$, it implies
\begin{align}
    0 = & \nabla \mathrm{LCB}(\mathbf{\tilde {z}}^* \vert  \mathcal{D}_r) + \sum_{i=1}^q \nabla\phi_g^i(\mathbf{\tilde {z}}^*) \nonumber \\
    = & \nabla \mathrm{LCB}(\mathbf{\tilde {z}}^* \vert  \mathcal{D}_r) -  c^{-1}\sum_{i=1}^q \frac{1}{\tilde g_i(\mathbf{\tilde {z}}^*)} \nabla \tilde g_i(\mathbf{\tilde {z}}^*) \label{eqn:dual}.
\end{align}
Then we introduce the dual variables $\lambda_i = -1/(c \tilde g_i(\mathbf{\tilde {z}}))$, the solutions $\mathbf{\tilde {z}}^*$ are also solutions to the following Lagrangian
\begin{equation}
    L(\mathbf{\tilde {z}}, \mathbf{\lambda}_i) = \mathrm{LCB}(\tilde{\mathbf{z}} \vert  \mathcal{D}_r) + \sum_{i=1}^q \lambda_i \tilde g_i(\mathbf{\tilde {z}}).
\end{equation}
Specifically, the dual function $l(\lambda_i)$ at $\mathbf{\tilde {z}}^*$ satisfies
\begin{equation*}
    l(\lambda_i^*) = \mathrm{LCB}(\tilde{\mathbf{z}}^* \vert  \mathcal{D}_r) + \sum_{i=1}^q \lambda_i^* \tilde g_i(\mathbf{\tilde {z}}^*) = \mathrm{LCB}(\tilde{\mathbf{z}}^* \vert  \mathcal{D}_r) - \frac{q}{c}.
\end{equation*}
The above equation quantifies the duality gap on any solution satisfying \eqref{eqn:dual}, i.e., any solution ultimately recommended by SB2O at each round. Therefore, we obtained the worst-case sub-optimality quantification in \eqref{eqn:prop_optimal}.
$\hfill\Box$



\begin{remark}
    \label{remark:safeBO}
    There are two basic safe BO frameworks in literature. The most popular one in machine learning ensures probabilistic safety with regret analysis \cite{BerkenkampKS16}. 
    To conduct safe exploration, additional conditions on Lipschitz continuity of both objective and constraints were needed in this framework, and the discretization of the search space $\Omega$ was also required in most of its extensions. In comparison, SB2O focuses optimization, mediating the exploration/exploitation balance by setting $\beta$ and $c$. In addition, since equation \eqref{eq:SB2O} is differentiable with analytical gradients, we do not deed discretization. The other safe BO framework also adopted the interior-point method, which was claimed to be compatible with any acquisition functions \cite{KrishnamoorthyD22}. Unfortunately, it lacks the theoretical analysis on both safety and sub-optimality. In comparison, employing LCB enables SB2O to have good characteristics for both theoretical and pragmatic perspectives.
\end{remark}

\begin{figure}[tb]
    \centering
    \scalebox{.52}{\includegraphics{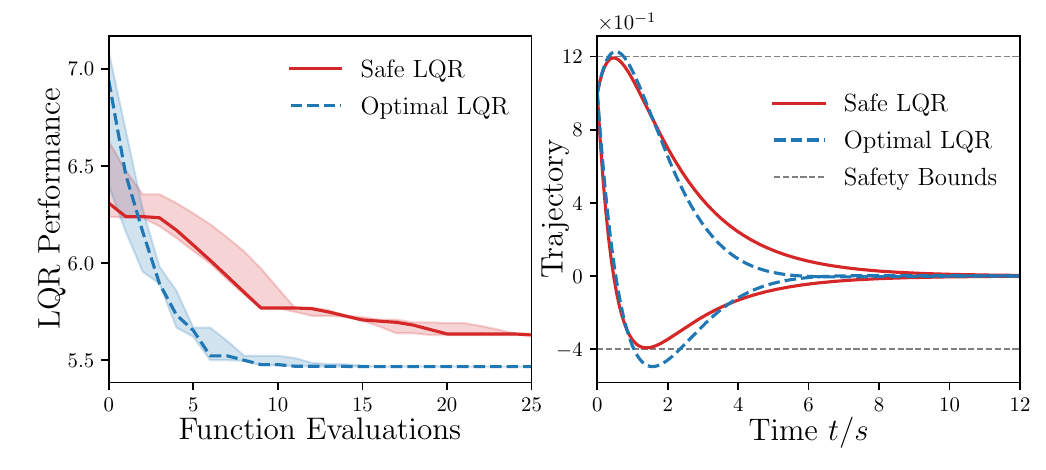}}
    \caption{An example of optimal control for linear systems via vanilla BO (optimal LQR) and SB2O (safe LQR). The search space is $[-5, -0.001]^2$. The best solution for optimal LQR is $[-1.0, -1.732]$ (same as the solution to the Riccati equation), and the one for safe LQR is $[-1.004, -2.324]$. (Left) The mean and variance of best observations during search with $5$ repetitive experiments. (Right) Trajectories of system state with different gains.}
    \label{fig:linearLQR}
\end{figure}


\emph{Illustrative Example 2}: Consider a second-order system $\ddot x = u$. Initially, $x(t_0) = \dot x(t_0) = 1$. The objective is to find the optimal control input gain $K=[k_1, k_2]$ such that a controller $u = K[x, \dot x]^\top$ yields the best LQR performance defined in \eqref{eqn:LQR_performance}. Let $Q=I$ and $R=I$. Conventionally, the optimal gain $K_r^\ast$ can be obtained by solving a well-defined Riccati function \cite{LewisVS12}. Alternately, we can conduct the optimal parameter adaptation with a few evaluations through BO. The results are shown in Fig. \ref{fig:linearLQR}, where the optimal gain can be found within $15$ evaluations. With a safety constraint such that $x$ and $\dot x$ should be bounded by $[-0.4, 1.2]$, conventional routines needs more complex computations \cite{LimonASC06}. In this situation, we can utilize the SB2O. The results are also presented in Fig. \ref{fig:linearLQR}. Likewise, the optimal parameter adaptation can be efficiently obtained within $30$ evaluations. As shown in the right panel of Fig. \ref{fig:linearLQR}, in addition to the good performance, this solution is safe regarding the state constraint, which is significant in SC2Ts.

\section{Experiments}
\label{sec:experiment}

We conduct two simulation experiments on swing-up control of a cart-pole system \cite{Tedrake22} and ACC for autonomous vehicles in CARLA simulator \cite{Dosovitskiy17}. All illustrative examples and experiments are performed on a desktop with Intel(R) Xeon(R) CPU E5-2620 v4 (2.10GHz) and NVIDIA GeForce GTX 1080Ti GPU. The GP models and CBO algorithm are implemented based on GPflow and GPflowOPT \cite{MatthewsWNFBLGH17}. Two comparative algorithms are considered. The first one is random search strategy (dubbed \texttt{CBF-RS}) to mimic manual adjustment, where LHS is used to uniformly sample the candidate solutions. The other algorithm is a powerful black-box optimization method, the covariance matrix adaptation evolution strategy (CMA-ES) dubbed \texttt{CBF-CMA} \cite{ArnoldH12}, which is implemented based on pycma \cite{hansen2019pycma}. 
To obtain acceptable solutions in limited evaluations, we set the population size as $2$ and initial deviation as $1$. 
Another safe BO framework discussed in Remark \ref{remark:safeBO} is not included in our experiments due to the impracticality of discritizing the searching space in SC2Ts \cite{BerkenkampKS16}. 
For all algorithms, $20$ repetitive experiments are conducted with different initialization.


\subsection{Swing-Up Control for Cart-Pole Systems}

\subsubsection{System Descriptions} 
\begin{figure}[tb]
    \centering
    \scalebox{.6}{\includegraphics{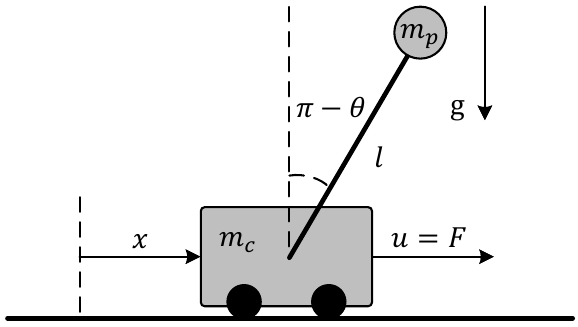}}
    \caption{The physical structure of a 2-D cart-pole system.}
    \label{fig:cartpole_system}
\end{figure}

The physical structure of a 2-D cart-pole system is depicted in Fig. \ref{fig:cartpole_system}, whose dynamics can be formulated by
\begin{align}
    \dot p = 
    \left[\begin{matrix}
    \dot x \\
    \dot v   \\
    \dot \theta\\
    \dot \omega
    \end{matrix}\right] =  \left[\begin{matrix}
    v\\
    f_{v}(\theta, \omega) \\
    \omega\\
    f_{\omega}(\theta, \omega)
    \end{matrix}\right] + \left[\begin{matrix}
    0 \\
    g_v(\theta)   \\
    0\\
    g_\omega(\theta)
    \end{matrix}\right] u, \label{eqn:cart_pole_system_dynamics}
\end{align}
where 
\begin{gather*}
    f_v(\theta, \omega) = \frac{m_p \sin \theta \left( l \omega^2 + g\cos \theta \right)}{m_c + m_p \sin^2{\theta}},\\
    f_\omega(\theta, \omega) = -\frac{m_p l \omega^2\cos \theta\sin\theta + \left( m_c + m_p \right)g\sin\theta}{l\left(m_c + m_p \sin^2{\theta}\right)}, \\
    g_v(\theta) = \frac{1}{m_c + m_p \sin^2{\theta}}, ~
    g_\omega (\theta) = -\frac{1}{l\left(m_c + m_p \sin^2{\theta}\right)}.
\end{gather*}

The objective of the swing-up control is to manipulate the pendulum, starting from the initial angle $\theta=0$ and cart position $x=x_0$, to swing above the cart such that $\theta$ reaches the desired angle $\theta_d = \left(2 n + 1\right) \pi$, where $n \in \mathbb{N}$. Simultaneously, the cart should be positioned at $x = x_d$. 
In what follows, we introduce the safe control framework of this task, and then optimize its configurable parameters. 
 
\subsubsection{Nominal controller} According to the energy shaping method \cite{Tedrake22}, we define the error of control energy as
\begin{align}
    \tilde{E} =  \frac{1}{2} m_p l^2 \omega^2 - m_p g l  -  m_p g l \cos{\theta}. \label{eqn:energy_shaping}
\end{align}
To ensure convergence of $\vert\tilde{E} \vert$, we introduce an additional proportional-differential (PD) controller and define $\ddot x_d$ as
\begin{equation}
    \ddot x_d = k_E \omega \cos{\theta}\tilde{E} - k_p x - k_d v,
\end{equation}
where $k_E>0$, $k_p\ge0$ and $k_d\ge 0$ are configurable parameters to be calibrated. The nominal controller in \eqref{eqn:cart_pole_system_dynamics} is designed by
\begin{equation}
    \begin{split}
        u_d = &\left(m_c + m_p - m_p\cos^2{\theta}\right) \ddot x_d \\
        & -  m_p g \sin{\theta} \cos{\theta} - m_p l \omega^2 \sin{\theta}.
    \end{split}
\end{equation}

To improve robustness, an LQR controller with feedback gain $K_{LQR}$ is switched on when system state approaches to the target state \cite{BruntonK19}. In all, given a switching threshold $\eta > 0$ and $p_d = [x_d, 0, \theta_d, 0]^\top$, we design the control input as
\begin{equation}
    u = \left\{\begin{matrix}
    u_d, & \text{initially},\\
    -K_{LQR} (p - p_d), & \text{once } \norm{p^{2-4} - p_d^{2-4}} \leq \eta.
    \end{matrix}\right.
\end{equation}

As a safety constraint, the position $x$ should be limited in $\left[-x_{l}, x_{l}\right]$. Accordingly, a CBF can be defined by $h_x = x_l^2 - x^2$, whose derivative is $\dot h_x = -2xv$. Since $\dot h_x$ does not explicitly contain control input $u$, an exponential control barrier function (ECBF) \cite{NguyenS16} is constructed by $h_x^e = \dot h_x + \mu h_x$ with $\mu>0$. For $\alpha>0$, the safety constraint derived from $h_x^e$ is
\begin{align}
    \dot h_x^e = & \ddot h_x + \mu \dot h_x \nonumber\\
    = & -2x\left(f_v(\theta, \omega) + g_v(\theta) u \right) - 2v^2 -2 \mu xv \nonumber\\
    \ge & -\alpha \left( -2xv + \mu\left( x_l^2 - x^2 \right) \right). \label{eqn:swingup_cbfconstraint}
\end{align}
Finally, the safe controller is derived by the following QP:
\begin{gather}
    \min_u  \frac{1}{2}(u - u_d)^2 \label{eqn:swingup_cbfqp_obj} \\
    s.t.~ \dot h_x^e(u) + \alpha \left( -2xv + \mu\left( x_l^2 - x^2 \right) \right)\ge 0. \tag{\ref{eqn:swingup_cbfqp_obj}{a}} \label{eqn:swingup_cbfqp_cbf}
\end{gather}

\subsubsection{Experiment settings} 
We consider two kinds of configurable parameters considered in this task, including the control parameters $k_E$, $k_p$, $k_d$, and safety parameters $\alpha$ and $\mu$, i.e.,  $\mathbf{z} = \left[k_E, k_p, k_d, \alpha, \mu\right]$. It is noted that the $K_{LQR}$ is calculated by first conducting local linearization and then solving the corresponding Riccati function.
In this task, we consider the LQR performance \eqref{eqn:LQR_performance} as the objective in \eqref{eq:cop}. Two constraints are considered. The first one is failure constraint considering the limitation in update frequency of QP solver. The other one is timeout constraint to save experiment costs.

\begin{figure}[t]
    \centering
    \scalebox{.529}{\includegraphics{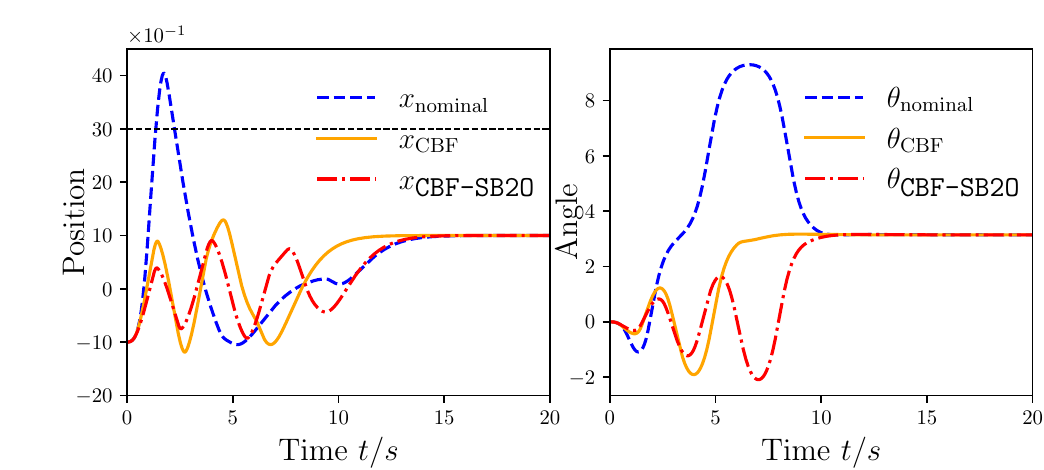}}
    \caption{Preliminary test on the swing up control using an unsafe nominal controller with $[k_E,k_p,k_d]=[0.3,0.8,1.0]$. We manually set the CBF parameters as $[\alpha,\mu] = [1.0, 1.0]$, with LQR performance $25.74$. After $40$ function evaluations by \texttt{CBF-SB2O}, we have $[\alpha,\mu] = [0.3, 7.7]$, with performance $22.44$. (Left) Trajectories of position of systems with different controllers. (Right) Trajectories of angle under different controllers.}
    \label{fig:cbf0-1}
\end{figure}

The cart-pole system is instantiated by $g=10.0\mathrm{m/s^2}$, $l=2.0\mathrm{m}$, $m_c=5\mathrm{kg}$, $m_p=1\mathrm{kg}$, the control input $u \in [ -100\mathrm{N}, 100\mathrm{N}]$. Initially, system is stationary at $x=-1.0\mathrm{m}$, $\theta=0 \mathrm{rad}$. The target position is $x_d=1.0\mathrm{m}$ and the maximal allowable position is $x_l = 3.0 \mathrm{m}$. Let $Q=I$ and $R=10^{-2}$ in LQR performance. Besides, the switch threshold is $\eta = 1.0$. 
A preliminary test is given in Fig. \ref{fig:cbf0-1} with manually-determined or optimized CBF parameters. The configuration parameters $\vartheta_f$ follow default settings in the QP toolbox \cite{DiamondB16}, where the update period of the filtered control input is fixed by $0.01\mathrm{s}$. The search space is $\Omega = \left[0.01,  100\right]^5$ (log scaled).
Each experiment of \texttt{CBF-SB2O} and \texttt{CBF-RS} contains $50$ initial solutions and $100$ function evaluations, while \texttt{CBF-CMA} takes the best solution found in initialization as the starting point.

\begin{figure}[t]
    \centering
    \scalebox{.532}{\includegraphics{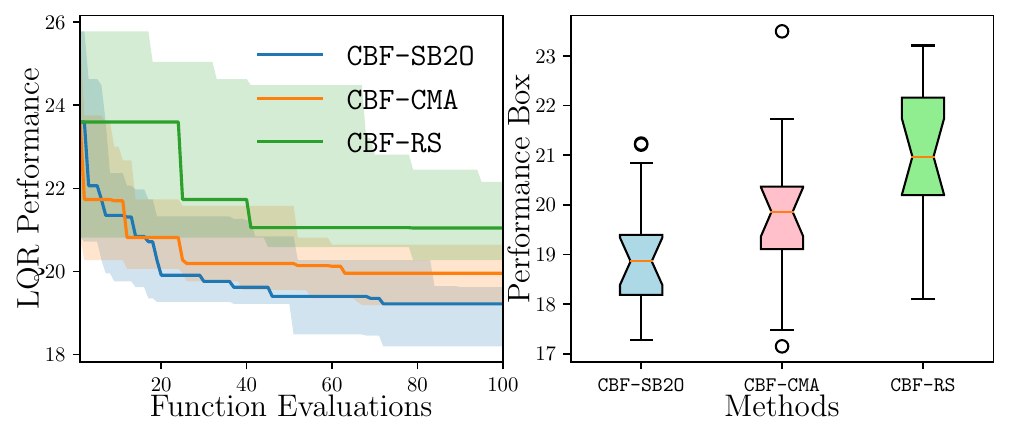}}
    \caption{The optimization result for swing up control via different algorithms. (Left) Performance optimization during $100$ function evaluations, where the median and $25/75$ percentiles of the best feasible objective value are plotted. (Right) Box plot of the final best performance of different algorithms. The median parameters found by \texttt{CBF-SB2O} is $[0.18, 0.03, 0.75, 0.56, 20.50]$.}
    \label{fig:swingup_result}
\end{figure}





\begin{table}[tb]
    \centering
    \caption{Numerical results of simulation experiments.}
    \begin{threeparttable}
    \begin{tabular}{c|c|cc}
    \hline
     Simulation & Method & Control Performance\tnote{a}  &  Feasibility Rate \\
     \hline
    \multirow{3}{*}{Swing up} & \texttt{CBF-RS} & $21.05 \;(0.04\%)$ & $54.92\%$  \\
     &\texttt{CBF-CMA} & $19.96 \;(0.81\%)$ & $89.13\%$  \\
     &\texttt{CBF-SB2O} & $\mathbf{19.22}^\dagger$ & $\mathbf{100}\%$  \\
     \hline
     
     \multirow{3}{*}{ACC} &\texttt{CBF-RS}& $200.11 \;(0.02\%)$ & $12.33\%$ \\
     &\texttt{CBF-CMA} & $194.31\;(2.07\%)$ & $67.25\%$  \\
     &\texttt{CBF-SB2O} & $\mathbf{189.91}^\dagger$ & $\mathbf{90.33}\%$  \\
     \hline
    \end{tabular}
    \begin{tablenotes}
            \item[a] The superscript $^\dagger$ indicates that the best algorithm is significantly better according to the Wilcoxon rank-sum test at $5\%$ significance level. The computing result is given in brackets.
        
    \end{tablenotes}
    \end{threeparttable}
    \label{tab:feasibility_reward}
\end{table}


\subsubsection{Experiment Results}
The simulation results are shown in Fig. \ref{fig:swingup_result}. Compared to a random safe configuration in Fig \ref{fig:cbf0-1}, the control performance is improved by all optimization methods. Differently, \texttt{CBF-SB2O} locates a better solution more efficiently throughout limited evaluations. As shown in Table \ref{tab:feasibility_reward}, \texttt{CBF-SB2O} is significantly better than other two algorithms. In terms of feasibility, it is shown that all configurations suggested by \texttt{CBF-SB2O} are feasible, while \texttt{CBF-RS} have a much higher probability to evaluate an infeasible solution. Similarly, the optimization efficiency of \texttt{CBF-CMA} is better than the random strategy, however, with more risks of conducting infeasible evaluations. 

\subsection{Adaptive Cruise Control for Autonomous Vehicles}


\begin{figure}[t]
    \centering
    \scalebox{.154}{\includegraphics{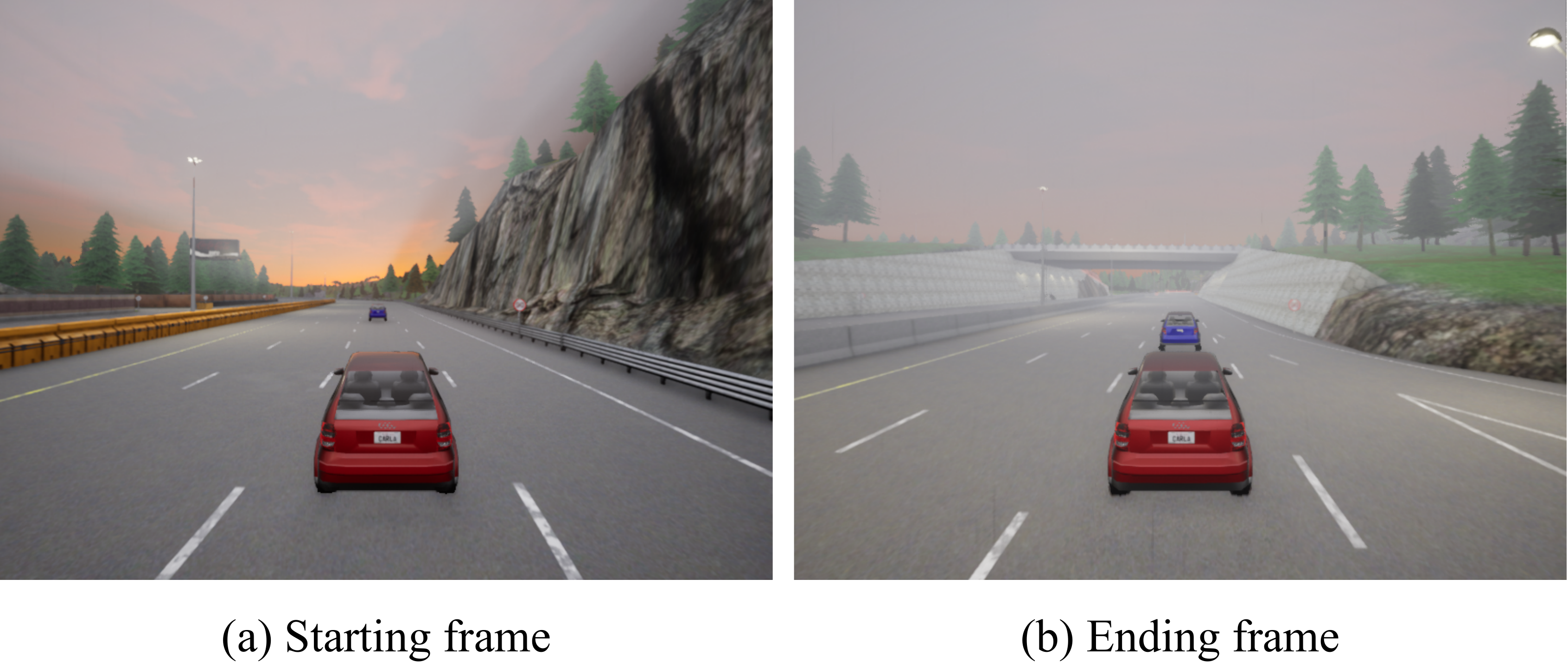}}
    \caption{Snapshots of ACC simulation in 'Town04'. The ego vehicle (red) should track the leader vehicle (blue) and keep a safe distance.}
    \label{fig:acc}
\end{figure}

Consider a leader vehicle (LV) starts and runs at a desired speed $v_0 = 64\mathrm{km/h}$, while an ego vehicle (EV), initially $31.4\mathrm{m}$ away from LF, should follow the leader while keeping a safe distance $d_{0} = 10\mathrm{m}$. The task environment is given in Fig. \ref{fig:acc}.
Note that in CARLA, the vehicle models are inaccessible. We assume no lane change happens and adopt PID controllers with fixed parameters for lane keeping (steering) \cite{LindemannRJDTM24}. For longitudinal control (throttle), we use the following dynamics:
\begin{equation}
    \dot v = -\frac{1}{m}F_r(v) + \frac{1}{m} u,\quad  \dot d = v_0 - v,
    \label{acc_vehicle_dynamics}
\end{equation}
where $v$, $d$ and $u \in [-0.4\mathrm{mg}, 0.4\mathrm{mg}]$ are the velocity of EV, distance between EV and LF, and EV throttle input, respectively. We use the blueprint \texttt{audi} vehicles for both EV and LV. We set parameters as $m = 1650\mathrm{kg}$, $f_0=2$, $f_1 = 5$, and $f_2 = 3$ (standard unit) combining manual adjustment and estimation techniques \cite{WangLWSZH23}, and fixed $g = 9.81 \mathrm{m/s^2}$.

Following the settings in \cite{XiaoB19}, the ACC controller consists of a CLF $V = (v-v_0)^2$ and a CBF $h = d_0 - d_0$. Given a slack variable $\delta $, we use QP with CBF\&CLF constraints and the following objective function 
\begin{equation}
    l_{\vartheta}(u,\delta) =\frac{1}{m^2} \left(u - F_r\right)^2 + k \delta^2,
\end{equation}
with $k > 0$. We also consider safety parameter $\alpha$ and QP computation frequency $f$ as the deployment parameter. Overall, the parameter to be optimized is $\mathbf{z} = [\alpha, k,  f]$. The objective function for ACC is designed as
\begin{equation}
    r = \int_{t_0}^{\infty} \frac{1}{v_0^2}( v - v_0)^2 + \frac{1}{d_0^2}(d-d_0)^2 + \frac{1}{m^2}(u-F_r)^2 \; dt.
\end{equation}

\begin{figure}[t]
    \centering
    \scalebox{.51}{\includegraphics{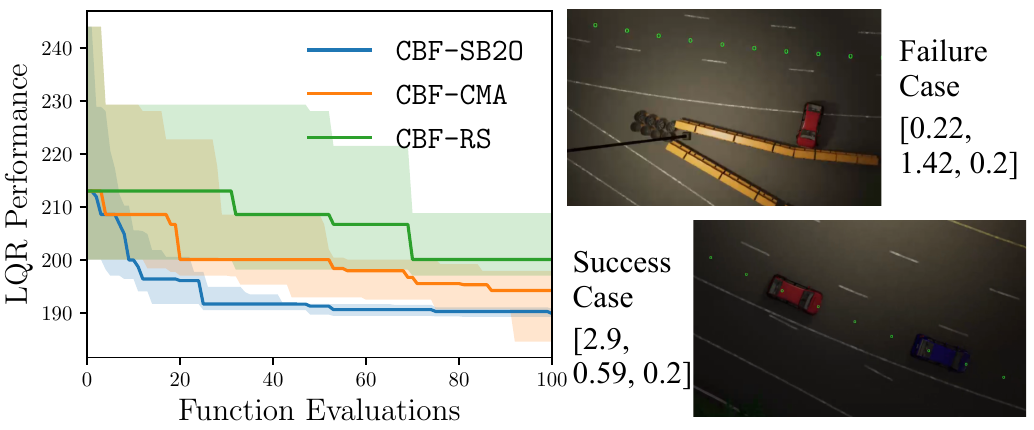}}
    \caption{(Left) Performance enhancement trajectories (median and $25/75$ percentiles) for ACC via different algorithms during $100$ function evaluations. The median parameters found by \texttt{CBF-SB2O} is $[ 3.51, 5.67, 0.1]$. (Right) Two successful and failed cases with the same $f$, where a smaller $\alpha$ is more likely to lead to the lane change.}
    \label{fig:acc_result}
\end{figure}

We use \texttt{Town04} as the ACC environment. The search space is $\Omega = [0.1, 10]^2\times[0.1, 1]$ (log scaled). The comparative results are presented in Fig. \ref{fig:acc_result} and Table \ref{tab:feasibility_reward}, where \texttt{CBF-SB2O} remains competitive against other two algorithms. Notably, the successful rate of \texttt{CBF-SB2O} is maintained beyond $90\%$, while other algorithms are more likely to fail due to model mismatches and noises in the simulator. The failure occurs in two scenarios. First, there is a chance that QP becomes infeasible during the ACC process. We also observed that $\alpha$ is more important than $f$ in influencing feasibility. The failure can also occur (with very small probability) when EV changed lane while turning due to the combination of \textit{fixed} PID-based steering and \textit{varying} CBF-based throttle input. This further highlights the importance of parameter adaptation in SC2Ts.

\section{Conclusions}
We proposed a systematic framework for optimal parameter adaptation of CBF-based controllers for SC2Ts. In consideration of the expensive black-box objective and constraints, a novel safe version of BO was developed for efficient and safe optimization. Two simulation experiments on swing-up control and high-fidelity adaptive cruise control demonstrated the competitiveness of \texttt{CBF-SB2O} in safe and efficient performance optimization of SC2Ts. 
One limitation of \texttt{CBF-SB2O} lies in that it can only work for episodic control tasks, making it not suitable for online parameter adaptation yet, which is likely to be solved by combining with DOP \cite{MaZTS22} or other approximation flow-based optimization techniques \cite{WangWYCSH23}.

\bibliographystyle{IEEEtran}
\bibliography{IEEEabrv,your_bib}

\end{document}